# Influence of Soil Plug on Pipe Ramming Process

## N. I. Aleksandrova

*Chinakal Institute of Mining, Siberian Branch, Russian Academy of Sciences,*
*Novosibirsk, 630091 Russia*
*e-mail: nialex@misd.ru*

**Abstract**—Different modes of advance of a pipe with a soil plug under rectangular impulse are investigated numerically and analytically with regard to dry friction between the pipe and plug and between the pipe and external stationary medium. The two model solutions are compared with and without regard to the pipe and plug elasticity. It is shown that elasticity of the pipe and plug is neglectable in case of a long-duration impulse.

## INTRODUCTION

One of the main objectives in trenchless laying of underground utilities by percussive pushing steel pipes in soil is the analysis of influence exerted on the wave process by friction between the pipe and soil inside and outside the pipe. Dry friction plays an important part in many mechanical systems with displacement of dry bodies. The issues of interaction between solids with regard to dry friction are addressed in many scientific publications [1–24]. The first to study the laws of dry friction were Leonardo da Vinci (1452–1519), Amonton (1663–1705) and Coulomb (1736–1806) [1]. A review of dry friction model can be found in [11, 19, 24]. Various aspects of numerical solution of problems connected with the nonlinear laws of dry friction were considered in [7–9, 15, 16, 21, 22, 24]. The analytical solutions for one degree of freedom systems with dry friction are given in [4, 6, 10, 13, 14, 17, 18, 20]. Stead-state motion of one or two bodies with dry friction described the Coulomb law [1] was analyzed in [2–4, 9, 12–14, 17], and nonsteady motion — in [9, 10, 18, 20–24]. Most of these studies are focused on the stick/slip change under harmonic driving force. Influence of impulse on the system of two bodies with dry friction remains yet to be investigated.

This study is focused on interaction between a cylindrical pipe, immobile external environment and a soil plug under the action of a single impulse. External and internal friction of the pipe is described by the classical Coulomb law [1].

## 1. ANALYTICAL SOLUTION FOR RIGID PIPE AND SOIL PLUG

We analyze joint motion of a steel cylindrical pipe in nondeformable soil and a soil plug inside the pipe under transient load $Q(t)$ directed along the pipe axis (Fig. 1). The pipe and soil plug are rigid, i.e. modeled by concentrated masses. The equations of motion with regard to the Coulomb law of constant dry friction at the inside pipe face and plug interface and at the outside pipe face and immobile soil interfaces (model I) have the form:

$$M_1 \ddot{U}_1(t) = Q(t) - P_1 L_1 \tau_1 k_1 - P_2 L_2 \tau_2 k_2, \qquad (1)$$

$$M_2 \ddot{U}_2(t) = P_2 L_2 \tau_2 k_2, \qquad (2)$$

$$k_1 = \operatorname{sign} \dot{U}_1 = \begin{cases} 1, & \dot{U}_1 > 0, \\ 0, & \dot{U}_1 = 0, \\ -1, & \dot{U}_1 < 0, \end{cases} \quad k_2 = \operatorname{sign}(\dot{U}_1 - \dot{U}_2) = \begin{cases} 1, & \dot{U}_1 - \dot{U}_2 > 0, \\ 0, & \dot{U}_1 - \dot{U}_2 = 0, \\ -1, & \dot{U}_1 - \dot{U}_2 < 0. \end{cases} \qquad (3)$$

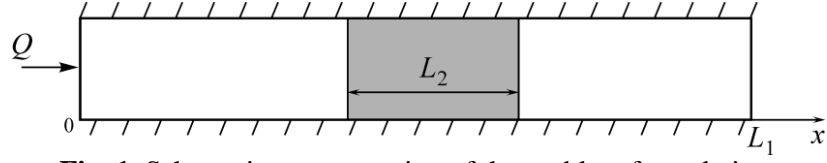

**Fig. 1.** Schematic representation of the problem formulation.

Here, $U_1$, $U_2$ are the displacements of the pipe and plug; $\tau_1$, $\tau_2$ are the limit shearing stresses at the outside and inside faces of the pipe; $P_1$, $P_2$ are the inner and outer perimeters of the pipe; $L_1$, $L_2$ are the pipe and plug lengths; $M_1$, $M_2$ are the pipe and plug masses; $t$ is the time. It is assumed in (1) and (2) that the dry friction force is applied over the whole outside pipe surface (proportional to $L_1$) and over the pipe and plug interface (proportional to $L_2$). It is hypothesized that all displacements and velocities are zero at the initial time $\dot{U}_1(0) = \dot{U}_2(0) = U_1(0) = U_2(0)$.

The further analysis is carried out in terms of the action of single rectangular impulse with the amplitude $Q_0$ and duration $t_0$:

$$Q(t) = Q_0 H(t) H(t_0 - t), \qquad (4)$$

where $H$ is the Heaviside step function.

Integrating of (1), (2) in view of (3), (4) and zero initial conditions yields:

$$\dot{U}_1(t) = \frac{1}{M_1}\left[ I(t) - \frac{Q_0 t(k_1 \tau_1 + \beta k_2 \tau_2)}{\tau_*} \right],$$

$$\dot{U}_2(t) = \frac{k_2 P_2 \tau_2 L_2}{M_2} t = \frac{Q_0 \tau_2 k_2}{M_2 \tau_*} \beta t \qquad (5)$$

where

$$\tau_* = \frac{Q_0}{P_1 L_1}, \quad \beta = \frac{P_2 L_2}{P_1 L_1}, \quad I(t) = \int_0^t Q(t) dt = Q_0[t H(t_0 - t) + t_0 H(t - t_0)].$$

The difference of the pipe and plug velocities is given by:

$$\dot{U}_1(t) - \dot{U}_2(t) = \frac{1}{M_1}\left[ I(t) - \frac{Q_0 t(k_1 \tau_1 + \gamma k_2 \tau_2)}{\tau_*} \right]$$

where $\gamma = \beta(1 + M_1/M_2)$.

In as much as the function $k_2 = 1$ when $\dot{U}_1 - \dot{U}_2 > 0$, we have the inequality to determine the time interval where $k_2 = 1$:

$$\dot{U}_1(t) - \dot{U}_2(t) = Q_0[t H(t_0 - t) + t_0 H(t - t_0) - (k_1 \tau_1 + \gamma \tau_2) t/\tau_*]/M_1 > 0. \qquad (6)$$

Similarly, we obtained the inequality to find the time interval with $k_1 = 1$:

$$\dot{U}_1(t) = Q_0[t H(t_0 - t) + t_0 H(t - t_0) - (\tau_1 + \beta k_2 \tau_2) t/\tau_*]/M_1 > 0$$

**a)** Let $\tau_1 + \gamma \tau_2 < \tau_*$. Then $\tau_1 + \beta \tau_2 < \tau_1 + \gamma \tau_2 < \tau_*$ as far as $\gamma > \beta$. The task is to find a time interval in which $k_2 = 1$ and $k_1 = 1$. Let $t_1$ and $t_2$ satisfy the equalities $t_1(\tau_1 + \gamma \tau_2) = \tau_* t_0$ and $t_2(\tau_1 + \beta \tau_2) = \tau_* t_0$. This implies $t_1 = \tau_* t_0/(\tau_1 + \gamma \tau_2) \geq t_0$ and $t_2 = \tau_* t_0/(\tau_1 + \beta \tau_2) > t_1 > t_0$. When $\tau_* = \tau_1 + \gamma \tau_2$, $t_1 = t_0$. In the time interval $t \in [0, t_1]$, we have $k_2 = 1$ and $k_1 = 1$. Consequently, the solution takes the form of:

$$\dot{U}_1(t) = \frac{Q_0}{M_1}[t H(t_0 - t) + t_0 H(t - t_0) - (\tau_1 + \beta \tau_2) t/\tau_*] H(t_1 - t),$$

$$\dot{U}_2(t) = \frac{Q_0 t \beta \tau_2}{M_2 \tau_*} H(t_1 - t).$$

The values of $\dot{U}_1$, $\dot{U}_2$ at the time $t_1$:

$$\dot{U}_2(t_1) = \frac{Q_0 t_0 \tau_2 \beta}{M_2(\tau_1 + \gamma\tau_2)}, \quad \dot{U}_1(t_1) = \frac{Q_0 t_0 \tau_2 \beta}{M_2(\tau_1 + \gamma\tau_2)}.$$

In this manner, until the time $t_1$, the pipe and plug have different velocities (the plug slips in the pipe) while at the time $t_1$ their velocities level (stick) and when $\tau_1/\tau_2 = \alpha$ ($\alpha$ = const), become the same:

$$\dot{U}_{**} = \frac{Q_0 t_0 \beta}{M_2(\gamma + \alpha)}$$

irrespective of the specific values of $\tau_1$, $\tau_2$.

Let $t > t_1$. Then inequality (6) is valid and, accordingly, $k_1 = 1$. The function $k_2$ at $t > t_1$ can assume two values: $k_2 = -1$ and $k_2 = 0$.

Suppose that $k_2 = -1$. From (5) we obtain the solution for the pipe in the time interval $t_1 < t < t_3$, where the time $t_3$ is found from the inequality $\dot{U}_1(t_3) = 0$:

$$\dot{U}_1(t) = \frac{Q_0(t_3 - t)(\tau_1 - \beta\tau_2)}{M_1 \tau_*} H(t_3 - t) H(t - t_1), \quad t_3 = t_1 \frac{\tau_1 + (\gamma - 2\beta)\tau_2}{(\tau_1 - \beta\tau_2)} \geq t_1.$$

In the time interval $t_1 < t < t_4$, where $t_4 = 2t_1$ is estimated from the equality $\dot{U}_2(t_4) = 0$, we have the solution for the soil plug velocity:

$$\dot{U}_2(t) = \frac{Q_0(t_4 - t)\beta\tau_2}{M_2 \tau_*} H(t - t_1) H(t_4 - t).$$

When $t > t_3, k_1 = 0$ and, consequently, $\dot{U}_1(t) = 0$; when $t > t_4$, $\dot{U}_2(t) = 0$.

Let us find the values of the parameters to satisfy the inequality $\dot{U}_1(t) - \dot{U}_2(t) < 0$ in the time interval $t_1 < t < t_3$, i.e. $k_2 = -1$. The inequality:

$$\dot{U}_1(t) - \dot{U}_2(t) = \frac{Q_0}{\tau_*}\left[\frac{(t_3 - t)(\tau_1 - \beta\tau_2)}{M_1} - \frac{(t_4 - t)\beta\tau_2}{M_2}\right] < 0$$

is only valid when $\gamma\tau_2 < \tau_1$. As a result, for $t > t_1$, we have $k_2 = -1$ in case that $\gamma\tau_2 < \tau_1$, otherwise $k_2 = 0$ if $\gamma\tau_2 \geq \tau_1$.

The ensuing solution at $\gamma\tau_2 < \tau_1$ and $\tau_1 + \gamma\tau_2 < \tau_*$ is given by:

$$\dot{U}_1(t) = \frac{1}{M_1}\left[I(t) - \frac{Q_0(\tau_1 + \beta\tau_2)t}{\tau_*}\right]H(t_1 - t) + \frac{Q_0(t_3 - t)(\tau_1 - \beta\tau_2)}{M_1 \tau_*} H(t_3 - t)H(t - t_1),$$

$$\dot{U}_2(t) = \frac{Q_0 \beta\tau_2}{M_2 \tau_*}[tH(t_1 - t) + (t_4 - t)H(t - t_1)H(t_4 - t)].$$

Let $t > t_1$ and $\gamma\tau_2 \geq \tau_1$. Then $k_2 = 0$ and $k_1 = 1$. In as much as the function $k_2 = 0$ in case that $\dot{U}_1(t) - \dot{U}_2(t) = 0$, then то $\dot{U}_1(t) = \dot{U}_2(t)$, i.e. the pipe and plug stick and move jointly. Their total mass is $M_1 + M_2$. The joint motion is described by the formula:

$$\dot{U}_1(t) = \dot{U}_2(t) = \frac{Q_0}{M_1 + M_2}\left(t_0 - \frac{\tau_1 t}{\tau_*}\right).$$

They move together until the time $t_5 = t_0 \tau_* / \tau_1$, if $\tau_1 \neq 0$:

$$\dot{U}_1(t) = \dot{U}_2(t) = \frac{Q_0(t_5 - t)\tau_1}{(M_1 + M_2)\tau_*} H(t_5 - t).$$

When $\tau_1 = 0$ then at $t > t_1$ we have $\dot{U}_1(t) = \dot{U}_2(t) = Q_0 t_0 /(M_1 + M_2)$.

Thus, when $\tau_1 + \gamma\tau_2 \leq \tau_*$ and $\gamma\tau_2 \geq \tau_1$, the solution is given by:

$$\dot{U}_1(t) = \frac{1}{M_1}\left[I(t) - \frac{Q_0 t(\tau_1 + \beta\tau_2)}{\tau_*}\right]H(t_1 - t) + \frac{Q_0(t_5 - t)\tau_1}{(M_1 + M_2)\tau_*}H(t - t_1)H(t_5 - t),$$

$$\dot{U}_2(t) = \frac{Q_0 t \tau_2}{M_2 \tau_*}H(t_1 - t) + \frac{Q_0(t_5 - t)\tau_1}{(M_1 + M_2)\tau_*}H(t - t_1)H(t_5 - t).$$

In case that $\tau_1 = \tau_2 = 0$, $\dot{U}_1(t) = I(t)/M_1$ and $\dot{U}_2(t) = 0$.

**b)** Let $\gamma\tau_2 + \tau_1 \geq \tau_*$. In the same way as above, there are two possible variants: $k_2 = -1$ and $k_2 = 0$.

It is supposed initially that $k_2 = -1$ and $k_1 = 1$. In this case, for $t \leq t_0$, it is valid that:

$$\dot{U}_1(t) = \frac{Q_0 t}{M_1}\left[1 - \frac{\tau_1 - \beta\tau_2}{\tau_*}\right]H(t_0 - t) > 0.$$

This inequality is fulfilled if $\tau_1 - \beta\tau_2 < \tau_*$.

Let us calculate the plug velocity in the interval $t \leq t_0$:

$$\dot{U}_2(t) = -\frac{Q_0 \beta t \tau_2}{M_2 \tau_*}H(t_0 - t)$$

and the relative pipe and plug velocity:

$$\dot{U}_1(t) - \dot{U}_2(t) = \frac{Q_0 t}{M_1}\left[1 + \frac{\gamma\tau_2 - \tau_1}{\tau_*}\right]H(t_0 - t).$$

The inequality $\dot{U}_1(t) - \dot{U}_2(t) < 0$ holds true when $\tau_1 - \gamma\tau_2 > \tau_*$.

On the plane $\tau_2$, $\tau_1$, the inequalities $\tau_1 - \gamma\tau_2 > \tau_*$, $\tau_1 - \beta\tau_2 < \tau_*$, $\tau_1 + \gamma\tau_2 \geq \tau_*$ define three domains with the single intersection point $(\tau_2, \tau_1) = (0, \tau_*)$. Finally, we have that $k_2 = -1$ and $k_1 = 1$ at $\tau_1 + \gamma\tau_2 \geq \tau_*$ and $t < t_0$ never exist.

Let $k_2 = 0$ and $k_1 = 1$. In this case, the pipe and soli plug move jointly:

$$\dot{U}_1(t) = \dot{U}_2(t) = \frac{Q_0 t}{M_1 + M_2}\left(1 - \frac{\tau_1}{\tau_*}\right)H(t_0 - t).$$

We calculate their velocities at the moment $t = t_0$:

$$\dot{U}_1(t_0) = \dot{U}_2(t_0) = \frac{Q_0 t_0}{M_1 + M_2}\left(1 - \frac{\tau_1}{\tau_*}\right).$$

Let $t > t_0$. We check the possibility of the situation when $k_2 = -1$. It is assumed that $k_2 = -1$. In this case, the inequality below should be fulfilled:

$$\dot{U}_1(t) = \frac{Q_0 t_0}{M_1 + M_2}\left[1 - \frac{\tau_1}{\tau_*} - \frac{\gamma(\tau_1 - \beta\tau_2)(t - t_0)}{(\gamma - \beta)\tau_* t_0}\right] > 0.$$

It is only valid when $\tau_1 < \tau_*$.

Let at $t = t_6$, $\dot{U}_1(t_6) = 0$. Then, we have:

$$t_6 = t_0\left(1 + \frac{(\tau_* - \tau_1)\gamma}{(\tau_1 - \beta\tau_2)(\gamma - \beta)}\right), \quad \dot{U}_1(t) = \frac{Q_0(t - t_6)(\beta\tau_2 - \tau_1)}{M_1 \tau_*}H(t_6 - t)H(t - t_0).$$

We calculate the plug velocity at $t > t_0$:

$$\dot{U}_2(t) = \dot{U}_2(t_0) - \frac{Q_0 \tau_2 \beta (t - t_0)}{M_2 \tau_*} = \frac{Q_0 t_0}{M_1 + M_2} \left(1 - \frac{\tau_1}{\tau_*} - \frac{\gamma \tau_2 (t - t_0)}{\tau_* t_0}\right) H(t - t_0)$$

and relative velocity of the pipe and plug:

$$\dot{U}_1(t) - \dot{U}_2(t) = \frac{Q_0 (\gamma \tau_2 - \tau_1)(t - t_0)}{M_1 \tau_*} H(t - t_0).$$

When $\gamma \tau_2 < \tau_1$, we have $\dot{U}_1(t) - \dot{U}_2(t) < 0$ and $k_2 = -1$ is realizable. Notice that for $\gamma \tau_2 < \tau_1$, $\beta \tau_2 < \tau_1$ and $t_6$ is determined.

Let $\dot{U}_2(t_7) = 0$ at $t = t_7$. We calculate this time:

$$t_7 = \frac{\gamma \tau_2 - \tau_1 + \tau_*}{\gamma \tau_2} t_0.$$

As a result, in case that the inequalities $\tau_1 + \gamma \tau_2 \geq \tau_*$ and $\gamma \tau_2 < \tau_1$ are satisfied, we have the plug velocity solution:

$$\dot{U}_2(t) = \frac{Q_0}{M_1 + M_2} \left[\left(1 - \frac{\tau_1}{\tau_*}\right) t H(t_0 - t) - \frac{\gamma \tau_2 (t - t_7)}{\tau_*} H(t - t_0) H(t_7 - t)\right].$$

Then, at $k_2 = 0$ ($\gamma \tau_2 \geq \tau_1$), the joint velocity of the pipe and plug is given by:

$$\dot{U}_1(t) = \dot{U}_2(t) = \dot{U}_1(t_0) - \frac{Q_0 \tau_1 (t - t_0)}{(M_1 + M_2) \tau_*} = \frac{Q_0}{M_1 + M_2} \left[t_0 - \frac{\tau_1}{\tau_*} t\right].$$

Here, $\dot{U}_1(t) > 0$ if $\tau_* > \tau_1$.

c) Let $\tau_1 > \tau_*$, then $\tau_1 + \beta \tau_2 > \tau_*$, $\tau_1 + \gamma \tau_2 > \tau_*$. Consequently, the pipe and plug remain motionless for the whole time interval:

$$\dot{U}_1(t) = \dot{U}_2(t) = 0, \quad t > 0.$$

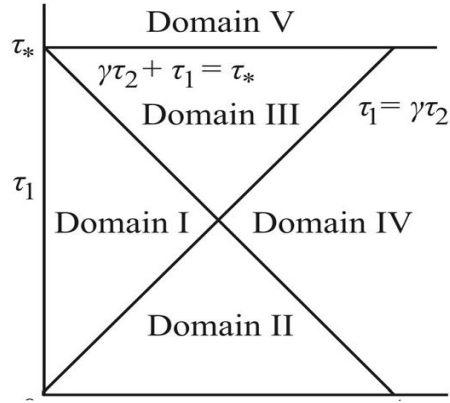

**Fig. 2.** Domains of solution in the plane $\tau_1$, $\tau_2$.

To sum up the results, we divide the plane $\tau_1$, $\tau_2$ into domains I, II, III, IV and V (Fig. 2). For each domain, we write expressions for the velocities of the pipe and plug and their displacements obtained from time integration of the pipe and plug velocities.

Solution in domain I ($\tau_2 \geq 0$, $\gamma \tau_2 + \tau_1 < \tau_*$, $\gamma \tau_2 < \tau_1$):

$$\dot{U}_1 = \frac{Q_0}{M_1 \tau_*} \left\{\left[\frac{I(t) \tau_*}{Q_0} - t(\beta \tau_2 + \tau_1)\right] H(t_1 - t) + (t - t_3)(\beta \tau_2 - \tau_1) H(t_3 - t) H(t - t_1)\right\}, \tag{7}$$

$$\dot{U}_2 = \frac{Q_0 \beta \tau_2}{M_2 \tau_*}[tH(t_1-t)+(t_4-t)H(t_4-t)H(t-t_1)],$$

$$U_1 = \frac{Q_0}{2M_1\tau_*}\{[\tau_* t^2 H(t_0-t)+\tau_*(2t-t_0)t_0 H(t-t_0)-(\tau_1+\beta\tau_2)t^2]H(t_1-t)$$

$$+[\tau_*(2t_1-t_0)t_0-(\tau_1+\beta\tau_2)t_1^2+(t-t_1)(t-t_1-2t_3)(\beta\tau_2-\tau_1)]H(t-t_1)H(t_3-t)$$

$$+[\tau_*(2t_1-t_0)t_0+t_3^2(\tau_1-\beta\tau_2)-2t_1^2\tau_1]H(t-t_3)\},$$

$$U_2 = \frac{Q_0\beta\tau_2}{2M_2\tau_*}[t^2 H(t_1-t)+(6t_1 t-4t_1^2-t^2)H(t_4-t)H(t-t_1)+4t_1^2 H(t-t_4)],$$

$$t_1 = \frac{\tau_* t_0}{\tau_1+\gamma\tau_2}, \quad t_3 = \frac{[(\gamma-2\beta)\tau_2+\tau_1]t_1}{\tau_1-\beta\tau_2}, \quad t_4 = 2t_1.$$

Solution in domain II ($\tau_1 \geq 0$, $\tau_1+\gamma\tau_2 < \tau_*$, $\gamma\tau_2 \geq \tau_1$):

$$\dot{U}_1 = \frac{Q_0}{M_1\tau_*}\left\{\left[\frac{I(t)\tau_*}{Q_0}-(\tau_1+\beta\tau_2)t\right]H(t_1-t)+\frac{\tau_1(\gamma-\beta)(t_5-t)}{\gamma}H(t_5-t)H(t-t_1)\right\}, \quad (8)$$

$$\dot{U}_2 = \frac{Q_0\beta\tau_2}{M_2\tau_*}\left\{tH(t_1-t)+\frac{\tau_1(t_5-t)}{\gamma\tau_2}H(t_5-t)H(t-t_1)\right\},$$

$$U_1 = \frac{Q_0}{2M_1\tau_*}\left\{[t^2\tau_* H(t_0-t)+(2t-t_0)t_0\tau_* H(t-t_0)-(\tau_1+\beta\tau_2)t^2]H(t_1-t)\right.$$

$$\left.+\left[(2t_1-t_0)t_0\tau_*-(\tau_1+\beta\tau_2)t_1^2+\frac{(\gamma-\beta)\tau_1(2t_5-t+t_1)(t-t_1)}{\gamma}\right]H(t_5-t)H(t-t_1)\right.$$

$$\left.+\left[(2t_1-t_0)t_0\tau_*-(\tau_1+\beta\tau_2)t_1^2+\frac{(\gamma-\beta)\tau_1(t_5^2-t_1^2)}{\gamma}\right]H(t-t_5)\right\},$$

$$U_2 = \frac{Q_0\beta\tau_2}{2M_2\tau_*}\left\{t^2 H(t_1-t)+\left[t_1^2+\frac{(2t_5-t+t_1)(t-t_1)\tau_1}{\gamma\tau_2}\right]H(t-t_1)H(t_5-t)\right.$$

$$\left.+\left[t_1^2+\frac{(t_5^2-t_1^2)\tau_1}{\gamma\tau_2}\right]H(t-t_5)\right\}, \quad t_5 = t_0\frac{\tau_*}{\tau_1}.$$

Solution in domain III ($0 \leq \tau_1 < \tau_*$, $\tau_1+\gamma\tau_2 \geq \tau_*$, $\gamma\tau_2 < \tau_1$):

$$\dot{U}_1 = \frac{Q_0}{(M_1+M_2)\tau_*}\left[t(\tau_*-\tau_1)H(t_0-t)+\frac{\gamma(t-t_6)(\beta\tau_2-\tau_1)}{(\gamma-\beta)}H(t_6-t)H(t-t_0)\right], \quad (9)$$

$$\dot{U}_2 = \frac{Q_0 t(\tau_*-\tau_1)}{(M_1+M_2)\tau_*}H(t_0-t)+\frac{Q_0(t_7-t)\beta\tau_2}{M_2\tau_*}H(t_7-t)H(t-t_0),$$

$$U_1 = \frac{Q_0}{2\tau_*(M_1+M_2)}\left\{t^2(\tau_*-\tau_1)H(t_0-t)\right.$$

$$\left.+\left[t_0^2(\tau_*-\tau_1)+\frac{\gamma(t-t_0)(t-t_0-2t_6)(\beta\tau_2-\tau_1)}{(\gamma-\beta)}\right]H(t_6-t)H(t-t_0)\right.$$

$$\left.+\left[t_0^2(\tau_*-\tau_1)+\frac{\gamma(t_6^2-t_0^2)(\tau_1-\beta\tau_2)}{(\gamma-\beta)}\right]H(t-t_6)\right\},$$

$$U_2 = \frac{Q_0 \beta}{2M_2 \tau_* \gamma}\{t^2(\tau_* - \tau_1)H(t_0 - t) + [t_0^2(\tau_* - \tau_1) + (t - t_0)(2t_7 - t + t_0)\gamma\tau_2]H(t_7 - t)H(t - t_0)$$
$$+ [t_0^2(\tau_* - \tau_1) + (t_7^2 - t_0^2)\gamma\tau_2]H(t - t_7)\},$$

$$t_6 = \left(1 + \frac{(\tau_* - \tau_1)(\gamma - \beta)}{(\tau_1 - \beta\tau_2)\gamma}\right)t_0, \quad t_7 = \frac{\gamma\tau_2 - \tau_1 + \tau_*}{\gamma\tau_2}t_0.$$

Solution in domain IV ($\tau_1 < \tau_*$, $\tau_1 + \gamma\tau_2 \geq \tau_*$, $\gamma\tau_2 \geq \tau_1$):

$$\dot{U}_1 = \dot{U}_2 = \frac{Q_0}{(M_1 + M_2)\tau_*}[t(\tau_* - \tau_1)H(t_0 - t) + \tau_1(t_5 - t)H(t_5 - t)H(t - t_0)], \tag{10}$$

$$U_1 = U_2 = \frac{Q_0}{2(M_1 + M_2)\tau_*}\{[t_0^2(\tau_* - \tau_1) + \tau_1(2t_5 - t + t_0)(t - t_0)]H(t_5 - t)H(t - t_0)$$

$$+ t^2(\tau_* - \tau_1)H(t_0 - t) + [t_0^2(\tau_* - \tau_1) + \tau_1(t_5^2 - t_0^2)]H(t - t_5)\}, \quad t_5 = t_0\frac{\tau_*}{\tau_1}.$$

Solution in domain V ($\tau_2 \geq 0$, $\tau_1 \geq \tau_*$): $\dot{U}_1 = \dot{U}_2 = 0$, $t \geq 0$.

## 2. ALGORITHM FOR SOLVING FINITE DIFFERENCE EQUATION WITH REGARD TO DRY FRICTION

The system of equations (1)–(3) with zero boundary conditions was solved in terms of a unit impact using the explicit finite difference scheme:

$$U_1^{n+1} - 2U_1^n + U_1^{n-1} = \tau^2(Q^n - P_2L_2\tau_2k_2 - P_1L_1\tau_1k_1)/M_1,$$
$$U_2^{n+1} - 2U_2^n + U_2^{n-1} = \tau^2 P_2L_2\tau_2k_2/M_2.$$

Here, $\tau$ is the step of the finite difference grid with respect to the time $t$; $U_1^n = U_1(\tau n)$, $U_2^n = U_2(\tau n)$ are the displacements of the pipe and plug at $t = \tau n$; $Q^n = Q(\tau n)$ is the amplitude of external impact at $t = \tau n$.

The algorithm of solving with dry friction is described below. As far as direction and force of friction are unknown beforehand, the pipe velocities are first calculated for two possible signs of $k_1$ ($k_1 > 0$ and $k_1 < 0$) at the assumption that $k_2 = 0$:

a) in the first case ($k_1 > 0$), a fictitious velocity is introduced: $\dot{U}_1^{0+} = (U_1^{0+} - U_1^n)/\tau$ where $U_1^{0+} = U_1^{n+1} - \tau^2 L_1\tau_1 P_1/M_1$;

b) in the second case ($k_1 < 0$) another fictitious velocity is introduced: $\dot{U}_1^{0-} = (U_1^{0-} - U_1^n)/\tau$, where $U_1^{0-} = U_1^{n+1} + \tau^2 L_1\tau_1 P_1/M_1$.

In both cases, $U_1^{n+1}$ is calculated from the difference equation for the pipe without regard to friction: $U_1^{n+1} = 2U_1^n - U_1^{n+1} + \tau^2 Q^n/M_1$.

The two possible situations in this case are:

1. If the velocities $\dot{U}_1^{0+}$ and $\dot{U}_1^{0-}$ have the same sign, the true value $U_1^{n+1}$ out of $U_1^{0+}$, $U_1^{0-}$ is chosen such $U_1^{0k_1}$ that can reach the minimum:

$$\text{abs}(\dot{U}_1^{0k_1}) = \min[\text{abs}(\dot{U}_1^{0-}), \text{abs}(\dot{U}_1^{0+})].$$

2. If the velocities $\dot{U}_1^{0+}$ and $\dot{U}_1^{0-}$ have different signs or one of them vanishes, then, based on the assumption of passive friction, the real pipe velocity equals zero.

Then, we calculate the pipe and plug velocities for two possible signs of $k_2$ ($k_2 > 0$ and $k_2 < 0$) at the assumption that the value of $k_1$ is chosen at the previous step:

a) in the first case ($k_2 > 0$) fictitious velocities are introduced: $\dot{U}_1^{+k_1} = (U_1^{+k_1} - U_1^n)/\tau$, $\dot{U}_2^+ = (U_2^+ - U_2^n)/\tau$, where $U_1^{+k_1} = U_1^{n+1} - \tau^2 (L_1 \tau_1 P_1 k_1 + L_2 \tau_2 P_2)/M_1$, $U_2^+ = U_2^{n+1} + \tau^2 L_2 \tau_2 P_2 / M_2$;

b) in the second case ($k_2 < 0$), the other fictitious velocities are introduced: $\dot{U}_1^{-k_1} = (U_1^{-k_1} - U_1^n)/\tau$, $\dot{U}_2^- = (U_2^- - U_2^n)/\tau$, where $U_1^{-k_1} = U_1^{n+1} - \tau^2 (L_1 \tau_1 P_1 k_1 - L_2 \tau_2 P_2)/M_1$, $U_2^- = U_2^{n+1} - \tau^2 L_2 \tau_2 P_2 / M_2$. The value $U_2^{n+1}$ is found from the difference equation for the plug without regard to friction: $U_2^{n+1} = 2U_2^n - U_2^{n+1}$.

In the same way as before, two situations are possible:

1. If the velocities $\dot{U}_1^{+k_1} - \dot{U}_2^+$ and $\dot{U}_1^{-k_1} - \dot{U}_2^-$ have the same sign, then the true values $U_1^{n+1}$, $U_2^{n+1}$ out of $U_1^{+k_1}$, $U_2^+$ and $U_1^{-k_1}$, $U_2^-$, are chosen as the pair $U_1^{k_2 k_1}$, $U_2^{k_2}$ which can reach the minimum:

$$\text{abs}(U_1^{k_2 k_1} - \dot{U}_2^{k_2}) = \min[\,\text{abs}(U_1^{+k_1} - \dot{U}_2^+), \text{abs}(U_1^{-k_1} - \dot{U}_2^-)\,].$$

2. If the velocities $\dot{U}_1^{+k_1} - \dot{U}_2^+$ and $\dot{U}_1^{-k_1} - \dot{U}_2^-$ have different sings, or one of them vanishes, then, based on the assumption of passive friction, the real relative velocity of the pipe and plug equals zero. The pipe and plug stick and move together, consequently, friction between them is absent.

Thus, the problem of calculating the times of transition of the pipe from standstill (relative standstill of the pipe and plug) to motion (relative motion of the pipe and plug), constituting the main difficulty in analytical solutions, reduces to the revealing of points where $\dot{U}_1^{0+}$ and $\dot{U}_1^{0-}$ ($\dot{U}_1^{+k_1} - \dot{U}_2^+$ and $\dot{U}_1^{-k_1} - \dot{U}_2^-$) have different signs, or one of them becomes zero. As far as the calculations unambiguously determine the value and direction of the friction force, under solving in each time level is the linear problem in which the friction force is already determined and inserted on the right-hand side of the equation.

### 3. GRAPHIC PRESENTATION OF NUMERICAL AND ANALYTICAL SOLUTIONS

Figures 3–6 show the velocity–time relationships obtained for the pipe and plug using the finite difference method and analytically from the formulas (7)–(10) at varied limit shearing stress coefficients $\tau_1, \tau_2$. The values of the other parameters of the problem are: pipe and plug lengths $L_1 = L_2 = 1$ m, pipe and plug densities $\rho_1 = 7800$ kg/m$^3$, $\rho_2 = 1800$ kg/m$^3$, pipe thickness $h_1 = 0.003$ m, Inner pipe radius $R_1 = 0.035$ m, impulse duration $t_0 = 50$ ms, effective force amplitude $Q_0 = 23\,000$ N, difference grid time step $\tau = 0.2$ ms. These values conformed with the pipe and plug masses $M_1 \approx 5.37$ kg, $M_2 \approx 6.93$ kg. The analytical and numerical solutions in Figs. 3–6 coincided up to the error of plotting; for this reason, the method by which the velocities are calculated is not mentioned in what follows. In Figs. 3–6, the solid lines show the functions $\dot{U}_1(t)$, the dash-and-dot lines — $\dot{U}_2(t)$ and the vertical dashed lines — the times $t_0, t_1, t_3, t_4, t_5, t_6, t_7$.

The velocities of pipe and plug in Fig. 3 are calculated from model I at $\tau_1 = \tau_* /(\gamma + 1) = 0.036564$ MPa. The problem parameters for curves *1* and *2* fit solution domain I; for curves *3*—domain IV. The analysis of the solutions at $\tau_1 = \tau_* /(\gamma + 1)$ shows that at $\tau_2 = 0$, the plug is immobile while the pipe is advanced; when $0 < \tau_2 < \tau_* /(\gamma + 1)$, the pipe and plug move at different velocity up to the time of arrestment; as $\tau_2 \geq \tau_* /(\gamma + 1)$ the pipe and plug stick and move together until stoppage.

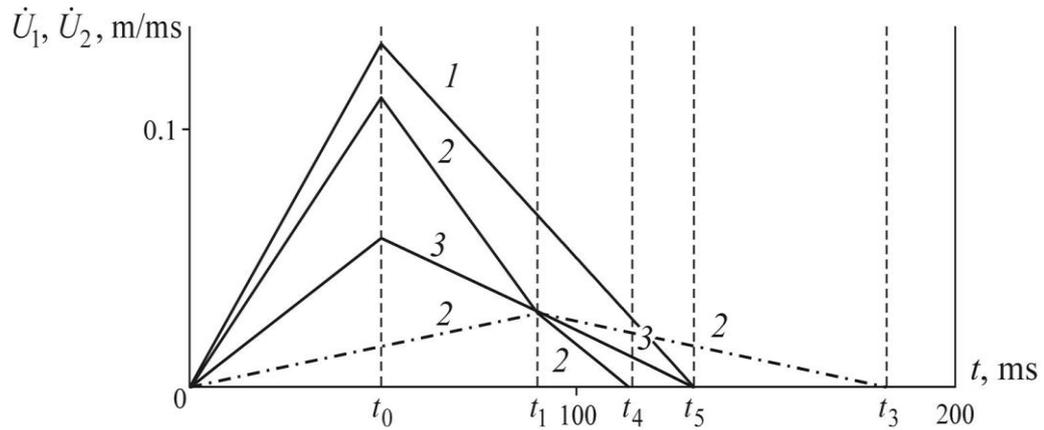

**Fig. 3.** Velocities of the pipe and plug at $\tau_1 = \tau_*/(\gamma+1)$: *1* — $\tau_2 = 0$; *2* — $\tau_2 = 0.01\,\text{MPa}$; *3* — $\tau_2 = 0.036564\,\text{MPa}$.

Figure 4 shows the pipe and plug velocities calculated from model I at $\tau_1 = 2\tau_2$. The problem parameters for curve *1* fit solution domain II, for curves *2*—domain I, for curves *3*, *4* — domain III. The horizontal dashed lines denote the values of $\dot{U}_* = Q_0 t_0 / M_1$ and $\dot{U}_{**} = Q_0 t_0 \beta / M_2 /(\gamma + \alpha)$ ($\alpha = 2$) Under parameters from domain III, the pipe and plug move jointly until the time $t_0$ and then have different velocity up to arrestment.

Figure 5 depicts the pipe and plug velocities calculated from model II at $\tau_1 = 0$. The problem parameters for curves *1–4* comply with domain II, for curve *5* — domain IV. The horizontal dashed line demonstrates the value of $\dot{U}_* = Q_0 t_0 /(M_1 + M_2)$. It is seen in the figure that without friction on the outer face of the pipe ($\tau_1 = 0$), the pipe and plug velocities gradually assume the constant value $\dot{U}_*$.

Figure 6a demonstrates the pipe and plug velocity curves from model I at $\tau_1 = \tau_2$. The problem parameters for curves *1–4* fit solution domain II, curves *5*, *6* — domain IV. The horizontal dashed lines show $\dot{U}_* = Q_0 t_0 / M_1$ and $\dot{U}_{**} = Q_0 t_0 \beta / M_2 /(\gamma + \alpha)$ ($\alpha = 1$). Apparently, as against domain I, in solution domain II, the pipe and plug first move at different velocities, which means slip, and then stick and move jointly. Furthermore, the value of $\dot{U}_{**}$ is constant at any limit shearing stresses at the inner and outer face of the pipe given that $\tau_1/\tau_2 = \text{const}$.

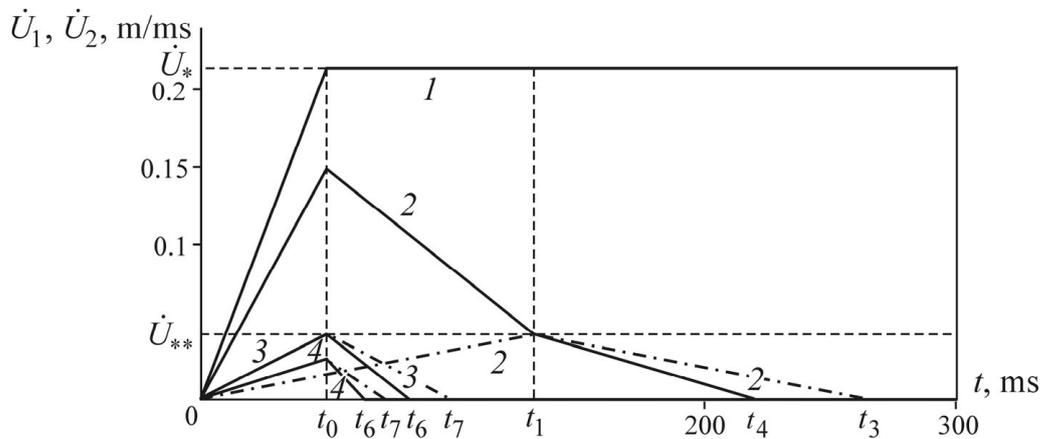

**Fig. 4.** Velocities of the pipe and plug at $\tau_1 = 2\tau_2$: *1* — $\tau_1 = 0$; *2* — $\tau_1 = 0.02\,\text{MPa}$; *3* — $\tau_1 = 0.053$; *4* — $\tau_1 = 0.07\,\text{MPa}$.

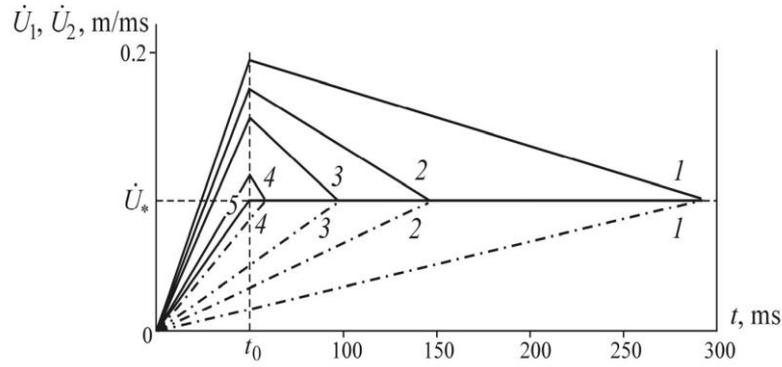

**Fig. 5.** Velocities of the pipe and plug at $\tau_1 = 0$: $1$ — $\tau_2 = 0.01$ MPa; $2$ — $\tau_2 = 0.02$; $3$ — $\tau_2 = 0.03$; $4$ — $\tau_2 = 0.05$; $5$ — $\tau_2 = 0.0589$ MPa.

Let the contact area between the pipe and outside soil is zero at the initial time. It is assumed that as pipe is driven in soil under external impact, the contact area between the pipe and soil grows in proportion to the velocity $U_t$. The equations of motion in this case (model II) are given by:

$$M_1 \ddot{U}_1(t) = Q(t) - P_1 U_1 \tau_1 k_1 - P_2 L_2 \tau_2 k_2, \quad M_2 \ddot{U}_2(t) = P_2 L_2 \tau_2 k_2. \tag{11}$$

It is difficult to solve (11) analytically and they were solved using the finite difference method therefore. Figure 6b shows the calculated results from model II at the same parameters as in Fig. 6a. The velocities are higher in Fig. 6b than in Fig. 6a., which can be explained by the fact that the friction force is lower in case it is proportional to the pipe penetration than in case it is proportional to the pipe length for a time interval.

## 4. COMPARISON OF THE SOLUTIONS OBTAINED WITH AND WITHOUT REGARD TO THE PIPE AND PLUG ELASTICITY

To understand the validity of modeling motion of pipe and soil plug as rigid concentrated masses, we analyze interaction between the elastic pipe and elastic plug. Their motion is described as the motion of elastic rods using one-dimensional wave equation in terms of displacements (model III):

$$\rho_1 S_1 \frac{\partial^2 u_1}{\partial t^2} = E_1 S_1 \frac{\partial^2 u_1}{\partial x^2} - P_2 \tau_2 \mathrm{sign}(\dot{u}_1 - \dot{u}_2), \tag{12}$$

$$\rho_2 S_2 \frac{\partial^2 u_2}{\partial t^2} = E_2 S_2 \frac{\partial^2 u_2}{\partial x^2} + P_2 \tau_2 \mathrm{sign}(\dot{u}_1 - \dot{u}_2). \tag{13}$$

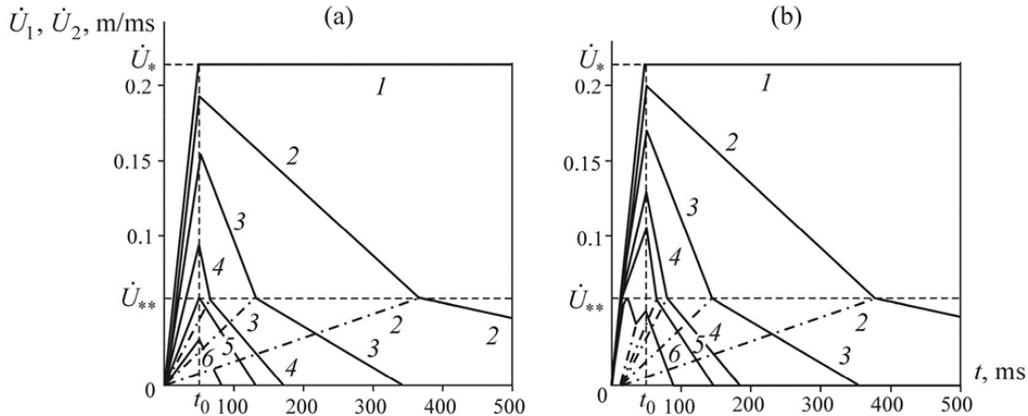

**Fig. 6.** Velocities of the pipe and plug at $\tau_1 = \tau_2$: (*a*) model I; (*б*) model III; $1$ — $\tau_2 = 0$; $2$ — $\tau_2 = 0.005$ MPa; $3$ — $\tau_2 = 0.014$; $4$ — $\tau_2 = 0.028$; $5$ — $\tau_2 = 0.036564$; $6$ — $\tau_2 = 0.065$ MPa.

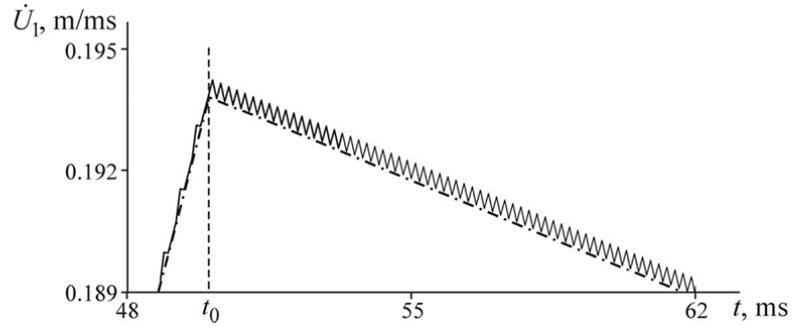

**Fig. 7.** Velocities of pipe: dash-and-dot line is model I; solid line is model III.

Here, $u_1$, $u_2$ are the displacements of the pipe and plug; $\tau_2$ is the limit complex stress between the pipe and plug; $P_2$ is the inner perimeter of the pipe; $S_1$, $S_2$ are the cross section areas of the pipe and plug; $E_1$, $E_2$ are Young's moduli of the pipe and plug; $\rho_1$, $\rho_2$ are the densities of the pipe and plug; $t$ is the time; $x$ is the axial coordinate. The initial conditions are zero. The coordinate system is chosen such that its origin coincides with the pipe end subjected to impacting and the axis $x$ is directed in parallel to the pipe axis (refer to Fig. 10). The longitudinal load $Q(t) = Q_0 H(t) H(t_0 - t)$ is applied at the cross section $x = 0$. At the ends of the pipe and plug, the boundary conditions are set:

$$E_1 S_1 \frac{\partial u_1}{\partial x}\bigg|_{x=0} = -Q(t), \quad E_1 S_1 \frac{\partial u_1}{\partial x}\bigg|_{x=L_1} = 0, \quad E_2 S_2 \frac{\partial u_2}{\partial x}\bigg|_{x=0} = 0, \quad E_2 S_2 \frac{\partial u_2}{\partial x}\bigg|_{x=L_2} = 0. \quad (14)$$

The problem was solved by the cross-type explicit finite difference scheme. The algorithm to calculate dry friction in the numerical solution for a rod-like pipe is described in [21, 22].

Figure 7 shows fragments of the pipe velocities. The solid line shows the velocity in the middle section of the pipe ($x = L_1 / 2$) in case of model III for elastic bodies. For the rigid masses in model I, the velocity is denoted by the dash-and-dot line. The full plot for model I is depicted in Fig. 5 by curve *1*. The problems parameters are: $E_1 = 195 \cdot 10^6$, $E_2 = 0.6 \cdot 10^6$ MPa, $\tau_1 = 0$, $\tau_2 = 0.01$ MPa, $\tau = 0.002$ ms, the difference grid step along the longitudinal coordinate is $h = 0.01$ m, the other parameters are the same as in Figs. 3–6.

It is seen in the figure that the behavior of the solutions in the elastic and rigid models of interaction is qualitatively the same. Differences are oscillations that arise in the elastic model solutions due to reflection of waves from the pipe ends. The error of the maximum pipe velocity amplitude in simpler model I is less than 0.25% relative to elastic model III. The comparison of the solutions obtained by the two models shows that in case when the impulse duration is much longer than the time of to and fro travel of wave along the pipe at a rod velocity, the pipe and plug motion can be described using the simplified model (1), (2).

## CONCLUSIONS

The analytical solutions to describe ramming of pipe with a soil plug are obtained. Various modes of the pipe and plug system motion are studied depending on the values of the limit shearing stresses at the outer and inner faces of the pipe. The finite difference algorithms are developed for different models of the pipe and plug with a view to dry friction. It is shown that the numerical and analytical solutions match together at a high accuracy. It is found that in case of a sufficiently long impulse, the influence of elasticity of the pie and plug can be neglected and the pipe and plug motion can be modeled as concentrated masses.

## ACKNOWLEDGMENTS

This study was supported by the Russian Science Foundation, project no. 17-77-20049.